# Unseen facts due to the statistical derivation of the Bell inequality and their logical consequences

Louis Sica[1,2]

[1]Institute for Quantum Studies, Chapman University, Orange, CA & Burtonsville, MD. USA

[2]Inspire Institute Inc., Alexandria, VA, USA

Email:lousica@jhu.edu

The Bell inequality is derived under the assumption of three physical data sets, random or deterministic. The data sets represent a laboratory realization of Bell's three probability based variables. For physical data as can be written on paper, the derivation of the inequality results only from principles of algebra and is independent of assumptions of locality, hidden variables and even randomness. Cross correlations of thee data sets carried out as Bell correlated three random variables results in the same inequality that is identically satisfied even by deterministic data. However, to obtain three data sets on two particles destroyed by measurement, two experimental runs are required, followed by data matching to reduce four data sets to three. If quantum mechanical probabilities are used to describe the data frequencies, the Bell inequality is satisfied. The situation is analogous to performing two sets of coin flips to compare the effect of two different coin loadings on the probabilities for heads and tails.

## 1. Introduction

The Bell theorem [1] and inequality, together with violation of the inequality under certain experimental procedures, have led to more than 50 years of speculation and controversy. [2,3]. The author some time ago [4] discovered a re-derivation of the Bell inequality applicable to three laboratory data sets independent of Bell's assumptions of locality or hidden variables. However, inconsistencies between the Bell derivation and requirements for experimental implementation require discussion. Below, Bell's inequality derivation is reviewed, and problems in its experimental realization in probability space are resolved by a data matching scheme that reduces four data sets to three, as required for inequality application. The intent of the present analysis is to provide a clearer understanding of the Bell inequality and theorem, and their implications.

## 2. Bell's inequality derivation from three cross-correlated variables

Bell [1] assumed three measurements applied to a pair of entangled spins or photons, as shown in the experimental schematic of Figure 1. He then derived a statistical expression using three variables applied to two particles, using observables seemingly accessible from the apparatus. However, only one measurement may be performed per particle, since measurements destroy the particles, and that has led to logical confusion. How is a three variable inequality, obtained as a mathematical construction, to be logically applied to the actual experimental results? More difficult yet, how is the four variable version of the inequality to be applied to such particle pairs?

First, the experimental situation must be presented based on the apparatus represented in Fig. 1. Measurements on the A-side at angular setting $a$ are represented in Bell's notation by the function $A(a,\lambda)$, and on the B-side at angular setting $b$ by $B(b,\lambda)$. The variables $\lambda$ are random variables with a probability density $\rho(\lambda)$ assumed to determine the results of measurements represented by functions $A$ and $B$, postulated to be deterministic. The random results of measurements might then be interpreted causally as due to uncontrolled initial conditions sampled randomly. The measurements have values equal to $\pm 1$ in Bell's definition with the additional stipulation that $B(a,\lambda) = -A(a,\lambda) = \pm 1$, so that the required measurement outcomes resulting from entanglement are fulfilled [1].

From the functions Bell defined, three mutual cross-correlations were computed between one measurement on the *A*-side and two measurements on the *B*-side of Figure 1. Using the Bell representation, the first of these cross-correlations is:

$$C(a,b) = \int A(a,\lambda)B(b,\lambda)\rho(\lambda)d\lambda \ . \tag{2.1}$$

(Bell used the variable *P* instead of *C* for the correlation, and this has occasionally led to misinterpreting the result as a probability rather than a correlation.) From this and similar expressions, Bell computed the absolute value of the difference of correlations for one observable on the left and two alternatives on the right of Fig. 1:

$$|C(a,b) - C(a,b')| = \left| \int A(a,\lambda)( B(b,\lambda) - B(b',\lambda) )\rho(\lambda)d\lambda \right|$$

$$\leq \int | A(a,\lambda)B(b,\lambda) || (1 - B(b,\lambda)B(b',\lambda) ) | \rho(\lambda)d\lambda$$

$$\leq \int |(1 - B(b,\lambda)B(b',\lambda) ) | \rho(\lambda)d\lambda$$

$$\leq \int |(1 - B(b,\lambda)B(b',\lambda) ) | \rho(\lambda)d\lambda = 1 - C(b,b') . \tag{2.2}$$

As noted, the fact that the inequality is based on three variables leads to an immediate difficulty. Two photons per realization emerge from the source in Figure 1. The detection of the photons destroys them. How then are three random variables to be obtained from an experiment that produces only one photon-pair per realization? This question is connected to an additional peculiarity. The correlations of (2.2) appear to have a mathematical symmetry. Each variable is correlated with two others. However, the physical situation indicated is intrinsically asymmetrical, due to the fact that two measurements must be obtained from one particle on the right-hand-side of Fig. 1 corresponding to a single measurement on the left-hand-particle, implying a different process of acquisition on the two sides.

Bell explicitly indicated (see Chap 8 of Bell's collected papers [5]) that the third random variable at $b'$ is defined simply as the outcome that would have occurred at that setting had it been used in place of $b$, given the same random variables $\lambda$. However, a result at only one or the other of the two *B*-side settings may be observed in the process of destroying the particle. Further, one cannot undo a random observation at $b$ that depends on random values of $\lambda$ to obtain another at $b'$. Even in strictly deterministic situations, an observation at one setting is not undone to compare it with the result at another setting. An additional measurement at the second setting is carried out and compared to that at the first.

*Thus, it must be concluded that the Bell inequality, if inextricably dependent on the notation used to derive it and Bell's prescription for its interpretation, does not precisely represent any experiment that may be performed using a single photon pair.*

To construct an operational response for Bell's prescription, consider the analogy of flipping a loaded coin or die. After a single flip, one may ask Bell's question: suppose the loading had been different on that flip, what would the result have been? This question must be answered in the probability domain, although the underlying process in this case is thought to be causal. The only experimental answer would be to change the loading and perform a large number of flips to determine the probability for heads and tails with the new loading. In the case of the common random process of coin flips, multiple interactions and ranges of causal variable values control the final outcome. Further, in such processes, a change in parameter in any decimal place may switch outcomes at a boundary between different outcomes. Thus, any implication that causality is equivalent to experimental predictability would seem to be unwarranted.

In the Bell experiment case, quantum mechanics provides predicted probabilities for a different instrument setting on a second particle, given a setting and outcome for the first. Thus, one can compare the quantum probability predictions with the experimental data although the procedure for acquiring this data must be carefully specified. The analogy above suggests an approach to applying inequality (2.2) to experimental data. As will be seen, the method for doing so is central to dealing with issues raised by the Bell theorem. However, a little known additional fact must first be considered.

## 3. A Bell inequality holds identically for three laboratory data sets

A pivotal physical fact relevant to the Bell theorem has yet to be noted. *In laboratory Bell experiments, correlations are not measured: finite data sets of ±1's are measured from which the correlations of (2.2) must be computed* [6]. *When the same algebraic steps are applied to the correlation estimates from finite data sets that Bell applied to correlations of infinite data sets, the same Bell inequality is obtained.* Given the important implications that follow from this result, it will be re-derived for examination below.

Assume that three data sets, random or deterministic, labeled $a$, $b$, and $b'$ have been obtained so that they can be written on paper. The data set items are denoted by $a_i$, $b_i$, and $b'_i$ with $N$ items in each set. Each datum equals $\pm 1$. One may begin by writing the equation

$$a_i b_i - a_i b'_i = a_i b_i (1 - a_i b_i a_i b'_i) , \qquad (3.1)$$

and sum it over the $N$ data triplets of the data sets. After dividing by $N$, one obtains

$$\frac{1}{N} \sum_{i=1}^{N} \left( a_i b_i - a_i b'_i \right) = \frac{1}{N} \sum_{i=1}^{N} a_i b_i (1 - a_i b_i a_i b'_i) . \qquad (3.2)$$

Taking absolute values of both sides,

$$\left| \frac{1}{N} \sum_{i=1}^{N} \left( a_i b_i - a_i b'_i \right) \right| = \left| \frac{1}{N} \sum_{i=1}^{N} a_i b_i (1 - a_i b_i a_i b'_i) \right| \le \frac{1}{N} \sum_{i=1}^{N} |a_i b_i| |(1 - a_i b_i a_i b'_i)| =$$

$$\frac{1}{N} \sum_{i=1}^{N} |(1 - a_i b_i a_i b'_i)|, \qquad (3.3)$$

or since $a_i^2 = 1$,

$$\left| \frac{\sum_i^N a_i b_i}{N} - \frac{\sum_i^N a_i b'_i}{N} \right| = \leq \frac{\sum_i^N |(1 - b_i b'_i)|}{N} = 1 - \frac{\sum_i^N b_i b'_i}{N}, \tag{3.4}$$

*The Bell inequality as applied to experimental data is thus a fact of algebra independent of the physical origin or properties of the data, and it holds for deterministic as well as random data.* The sums on the two sides of (3.4) have the form of correlation estimates although the data may be random, deterministic, or a combination of the two. In the case where the data are all random, they may exhibit correlations due to a variety of circumstances, e.g., the correlations may result from correlation to other variables not indicated or known. The final correlation of (3.4) reuses the data used to compute correlations of data sets (*a, b*) and (*a, b'*), and is not the result of new measurements. Thus, only three data sets are assumed, one for each of Bell's three cross-correlated variables. It is the cross-correlation of three variables, each equal to ±1, that produces the inequality with the constant 1 on the right-hand-side.

To repeat: No particular physical attributes have been necessary to obtain (3.4). The data sets may represent nonsense or be severely corrupted due to nonlocal interference between detectors. A startling conclusion follows: neither data fluctuations nor nonlocal interaction between detectors can cause violation of the Bell inequality under the condition of cross-correlation of *three physical data sets corresponding to the three cross-correlated variables assumed in Bell's derivation.*

In the case where the data are random and estimates statistically converge to correlations in (3.4) as *N* becomes large, one has for the first correlation $C_1(a,b)$ of variables *a* and *b*,

$$C_1(a,b) = \lim_{N \to \infty} \frac{\sum_i^N a_i b_i}{N},$$

with similar results for the other correlations of (3.4). The Bell inequality in correlations follows in the absence of Bell's various assumptions regarding the nature of the processes involved:

$$|C_1(a,b) - C_2(a,b')| \leq 1 - C_3(b,b'). \tag{3.5}$$

The correlation arguments $a, b$, etc., now refer to instrument angular settings while when subscripts are added, as in (3.4), they indicate individual data outputs at those settings. The three correlation functions will in general have different functional forms, as indicated by their subscripts, but without violating (3.5). However, it is critical to realize that the final correlational forms are constrained by (3.5). This follows from the fact that the two *B*-side data sets from two angular settings, corresponding to one setting on the *A*-side, are now themselves correlated. While the Bell inequality results from Bell's statistical derivation and assumptions regarding locality and a single probability density describing three variables, it clearly holds much more generally independently of these assumptions. *Bell's derivation is thus based on sufficient conditions. However, as results from (3.5), these conditions are unnecessary, and the inequality holds without them.*

## 4. Bell inequality violation from non-cross-correlated data

The Bell inequality is widely believed to be violated by carefully recorded experimental data. How can this be, given that the inequality is found to be identically satisfied by any three cross-correlated, physically obtained data sets independently of whether they are deterministic or random? Note that from the derivation reproduced in Sec. 2 it is not immediately obvious that the result of Sec 3 is also true. In the derivation of Sec. 2 various physical attributes of the data were originally spelled out such as independence of data at *A* from settings at *B, etc.* As seen above, this has no effect on the satisfaction of the inequality though it would certainly in general change the form of the correlations. Unfortunately, it has not been generally recognized that the constant in the inequality results from the cross-correlation of three variables (or four resulting in a different constant in the four variable case) and it is commonly believed that the correlations may be recorded using independent pairs of data [6]. The basic conditions that produce the constant in the two main versions of the Bell inequality are thus violated.

An important additional complication resides in the fact that under Bell's prescription, two (not three) measurement pairs are required to obtain the data for three correlations. Two data sets on the *A*-side must then be obtained in an actual experimental procedure so that three data sets may be created from two procedures. Bell himself did not recognize that data for the correlations $C(a,b)$ and $C(a,b')$ determined the correlation $C(b,b')$. Further, he thought that all three correlations had the same form, perhaps because two of the three did. This was based, however on a flip of the final output at variable $b'$ to its negative value on the opposite side of the apparatus at an equal setting [1]: $a'=b'$. Apparently this was thought to justify the conclusion that all the correlations were now of the same form.

Finally, evaluation of (3.4) should be briefly re-considered under the condition that variable pairs are obtained independently, as occurs in practice. Four data-sets are acquired for each pair of settings on the right-hand-side: $a_{1i}, a_{2i}, b_{1i}$, and $b'_{2i}$ (they occur in physical pairs and subscripts 1 and 2 indicate the different experimental runs). For a simple illustration of the consequences of usual practice, assume that all items of each set have a single value, that the two pairs are acquired independently, and variables have values $a_{1i}=1, a_{2i}=-1$, and $b_i = b'_i = 1$. One easily obtains violation of inequality (3.4) as achieved in Bell experiments:

$$|(1)(1)-(-1)(1)|=2 \neq 1-(1)(1)=0 \qquad (4)$$

## 5. Two quantum mechanical correlations determine the third

Note that the variables $B(b,\lambda)$ and $B(b',\lambda)$, given that two experimental runs are necessary to observe the Bell correlations needed for application of the inequality, are each correlated with $A(a,\lambda)$, and as a result are correlated with each other. In the derivation of (2.2) the relation $A(a,\lambda_i)B(b,\lambda_i)A(a,\lambda_i)B(b',\lambda_i) = B(b,\lambda_i)B(b',\lambda_i)$ has been used. If the correlation is computed by averaging over $\lambda$ in two statistically independent trials:

$$C(A(a,\lambda)B(b,\lambda)A(a,\lambda)B(b',\lambda)) = C(A(a,\lambda_1)B(b,\lambda_1))C(A(a,\lambda_2)B(b',\lambda_2)), \qquad (5a)$$

where subscripts 1 and 2 on $\lambda$ indicate the separate trials. The correlation then necessarily factors since probabilities in independent trials factor and must be multiplied. But in the Bell inequality derivation, the data must be selected so that $A(a,\lambda_1) = A(a,\lambda_2)$ since the same value of $A(a,\lambda)$ multiplies $B(b,\lambda)$ and $B(b',\lambda)$. It follows that

$$C(A(a,\lambda_1)B(b,\lambda_1))C(A(a,\lambda_2)B(b',\lambda_2)) = C(B(b,\lambda_1)B(b',\lambda_2)). \tag{5b}$$

Since each of the correlations on the left is given by the well-known Bell correlation, the result is that
$$C(B(b,\lambda_1)B(b',\lambda_2)) = [(-\cos(b-a))(-\cos(b'-a))] = \cos(b-a)\cos(b'-a) \tag{5c}$$

This important result will be re-derived below using quantum probabilities. As will be seen, $B(b,\lambda)$ and $B(b',\lambda)$ are variables each correlated to fixed outcomes at *A*, and so are correlated to each other.

The correlation of (5c) will now be explicitly derived using quantum mechanical probabilities to predict correlations at alternate variable setting pairs (*a*,*b*) and (*a*,*b'*) on the two sides of the apparatus of Fig. 1. These result from entanglement and are well known [7] (the subscripted pluses and minuses indicate ±1 outputs at instrument settings *a* and *b* respectively):

$$P_{++}(a,b) = P_{--}(a,b) = \tfrac{1}{2}\sin^2\frac{b-a}{2}; \quad P_{+-}(a,b) = P_{-+}(a,b) = \tfrac{1}{2}\cos^2\frac{b-a}{2}. \tag{5.1a}$$

The angular setting difference divided by 2 holds for Bell's original case of entangled spins. (In the optical version that corresponds to most Bell experiments, the 2 does not occur, with the result that a factor of 2 occurs in the argument of the final correlation.) Note, the joint probabilities (5.1a) are expressed in terms of conditional probabilities. Again using ± subscripts on the probabilities to indicate ±1 outcomes, $P_+(a) = P_-(a) = 1/2$ and the conditional probabilities of outcomes on the B-side given those on the A-side from (5.1a) are

$$P_{++}(b|a) = P_{--}(b|a) = \sin^2\frac{b-a}{2}; \quad P_{+-}(b|a) = P_{-+}(b|a) = \cos^2\frac{b-a}{2}. \tag{5.1b}$$

The probabilities at an alternative setting *b'* are obtained by inserting it in place of *b* in (5.1a). From these joint probabilities, the correlation $C(a,b)$ is

$$C(a,b) = [(+1)(+1)+(-1)(-1)]\tfrac{1}{2}\sin^2\frac{b-a}{2} + [(+1)(-1)+(-1)(+1)]\tfrac{1}{2}\cos^2\frac{b-a}{2}$$
$$= -\left(\cos^2\frac{b-a}{2} - \sin^2\frac{b-a}{2}\right) = -\cos(b-a), \tag{5.2a}$$

and the correlation $C(a,b')$ is immediately

$$C(a,b') = -\cos(b'-a). \tag{5.2b}$$

To implement Bell's prescription of Sec. 2 for two variables on the *B*-side of Fig. 1, two sets of observations must be performed just as in the case for observing probabilities in the analogous situation of two differently loaded coins. To be relatable to observations in Bell experiments, the probability densities used in the Bell notation must be replaced by quantum probabilities as appropriate to an ensemble of observations. The conditional probabilities of (5.1b) for setting coordinates (*b, b'*) on the same side of a Bell apparatus are then required. [8,9]

From (3.3,4) and corresponding steps in (2.2), given the physical situation, the correlation of outcomes at (*b,b'*) is the sum of conditional averages for $A_+(a)=1$ and $A_-(a)=-1$ each occurring with probability ½. In each case, the value observed at setting *a* is a parameter for the conditional probabilities now used to provide the products of probabilities for outcomes in the two independent trials. The normalization of these probabilities equals 1 for $A_+(a)=1$:

$$P_{++}(b|a)P_{++}(b'|a) + P_{-+}(b|a)P_{-+}(b'|a) + P_{++}(b|a)P_{-+}(b'|a) + P_{-+}(b|a)P_{++}(b'|a) =$$
$$\sin^2[(b-a)/2]\sin^2[(b'-a)/2] + \cos^2[(b-a)/2]\cos^2[(b'-a)/2] + \quad (5.3a)$$
$$\sin^2[(b-a)/2]\cos^2[(b'-a)/2] + \cos^2[(b-a)/2]\sin^2[(b'-a)/2] = 1,$$

A similar normalization may be computed for the opposite outcome at $A_-(a)=-1$.

Using the conditional probabilities from (5.1b), the conditional correlation $C(bb'|a,1)$ (in |a,1⟩, the 1 denotes the numerical output at setting *a*) is:

$$C(bb'|a,1) = (1)(1)\sin^2[(b-a)/2]\sin^2[(b'-a)/2] + (-1)(-1)\cos^2[(b-a)/2]\cos^2[(b'-a)/2] +$$
$$(1)(-1)\sin^2[(b-a)/2]\cos^2[(b'-a)/2] + (-1)(1)\cos^2[(b-a)/2]\sin^2[(b'-a)/2] = \quad (5.3b)$$
$$\cos(b'-a)\cos(b-a) \ .$$

Similarly:

$$C(bb'|a,-1) = \cos(b-a)\cos(b'-a) \ . \quad (5.3c)$$

Since the two values at *a* occur with probability ½ the overall correlation is

$$C(b,b'|a) = (1/2)C(bb'|a,1) + (1/2)C(bb'|a,-1) =$$
$$\cos(b-a)\cos(b'-a) \ . \quad (5.4)$$

After using appropriate trig identities, this has been shown to satisfy the Bell inequality (2.2) [10].

## 6. The Bell correlation is not unique to entanglement

It has been shown above that the Bell inequality holds independently of the assumption of locality, and the particular representation of random hidden variables commonly thought necessary to construct it. It follows from basic algebra applied to any three laboratory data sets (four in the four variable case) that can be written on paper. That fact does not in itself however, imply that random processes different from entanglement can produce Bell correlations. Nevertheless, a number of researchers have claimed demonstration of such a result based on computer algorithms, e.g., [11]. (Bell correlations have also been computed based on very small information transfer from detector *A* to detector *B* [12].)

These derivations are not physics-based, i.e., resulting from specific physical phenomena and properties that are intrinsically complex. Present physical models are effectively limited by an incomplete understanding of photons and their relation to electromagnetic waves. Hypothetical physical accounting for Bell correlations assuming that light consists of both waves and particles has nevertheless been given [13,14].

To show that the Bell correlation may result from a computer algorithm different from that implied by the Bell formalism, and that the correlation is not uniquely produced by entanglement, an example will be given. It will be developed from two independent detector settings, two independent Gaussian random variables, and a third random Poisson process used to determine whether a two-photon event occurs. The latter is reminiscent of a spontaneous two-photon down-conversion process as occurs in Bell-experiment sources [15]. The model developed was suggested by an example in a Papoulis monograph [16] that begins with a non-stationary random process in two continuous angle variables and imposes a condition that makes the correlation stationary. The version below is altered to produce pairs of $\pm 1 s$ rather than continuous variables.

Assume two functions using arbitrarily chosen settings $\theta_1$ and $\theta_2$:

$$z_1 = a\cos\theta_1 + b\sin\theta_1$$
$$z_2 = -a\cos\theta_2 - b\sin\theta_2 \tag{6.1}$$

with product

$$z_1 z_2 = -[a^2 \cos\theta_1 \cos\theta_2 + b^2 \sin\theta_1 \sin\theta_2 + ab(\cos_1 \sin\theta_2 + \sin\theta_1 \cos\theta_2)]. \tag{6.2}$$

If $a$ and $b$ are independent random variables with the same probability densities and zero mean, the average of (6.2) is

$$\overline{z_1 z_2} = -\overline{a^2}\cos\theta_1 \cos\theta_2 - \overline{b^2}\sin\theta_1 \sin\theta_2$$
$$= -\overline{a^2}\cos(\theta_1 - \theta_2), \tag{6.3}$$

where it will be assumed for use below that $\overline{a^2} = \overline{b^2} \ll 1$.

This construction needs to be converted to one with output variables equal to $\pm 1$. This is accomplished by dividing $z_1$ and $z_2$ by their absolute values $|z_1|$ and $|z_2|$:

$$\frac{z_1}{|z_1|}\frac{z_2}{|z_2|} = \frac{-(a\cos\theta_1 + b\sin\theta_1)(a\cos\theta_2 + b\sin\theta_2)}{|z_1||z_2|} \tag{6.4}$$

so that the factors being multiplied now equal $\pm 1$. But the average of (6.4) is not equal to that of (6.3). To accomplish this last step an uncertainty in count-pair production is introduced by assuming a Poisson-like process defined by the probability of a two-photon event occurrence $P(1) = |z_2||z_2| \ll 1$. (Probabilities with different arguments refer to different probability density functions.) The probability of a zero or non-event is given by $P(0) = 1 - P(1)$. The average value of the product (6.4) is then its value averaged by its rate of occurrence:

$$\sum P(a,b)[P(1||z_1||z_2|)1 + 0 P(0||z_1||z_2|)]\frac{z_1}{|z_1|}\frac{z_2}{|z_2|}$$
$$= \sum P(a,b)[|z_1||z_2|]\frac{z_1}{|z_1|}\frac{z_2}{|z_2|} = \overline{z_1 z_2} = -\overline{a^2}\cos(\theta_1 - \theta_2) \tag{6.5}$$

In this model, the values of $\theta_1$ and $\theta_2$ may be freely chosen. The derivation uses the probability that in some observation-time windows, no twin detections occur, as is the case in practice. Thus, the possibility that $A$ and $B$ may have the value 0 in addition to $\pm 1$ plays an important role.

## 7. Conclusion

It has been shown above that the statistical inequality in cross-correlations of three random variables derived by Bell, using an assumed representation of hidden variables, is duplicated if the same algebraic construction is applied to any three physically acquired laboratory data sets, random or deterministic. The inequality applied to physical data does not depend on the assumption of locality or on any particular representation of hidden variables as widely believed necessary for its derivation. For intrinsically finite laboratory data, expressions in the form of correlation estimates take the place of Bell's correlations.

In the random data case, fluctuation of the estimates cannot result in violation of the inequality, since it must be satisfied given the mere existence of three data sets, random or deterministic. (The same conclusion is easily demonstrated for the four variable inequality.) Thus, the inequality has very different implications than those thought to arise from Bell's formulation since none of the assumptions thought necessary to its derivation are in fact used.

The inequality is identically satisfied as a fact of algebra and an accounting for its violation in use is easily understood. The fact that it employs three variables as Bell prescribed to examine correlations of random measurements on particles only produced in pairs, has led to operational difficulties. The most important difficulty arises from the neglect to cross-correlate three observables, since it is the algebra involved in doing so that results in the inequality. Given that particles are produced in pairs in the physical situation to which the inequality has been applied, correlations of data pairs independently produced have been inserted into it without consideration of the inconsistency with inequality derivation. Since its use is at variance with the derivation, the inequality is violated. When a data acquisition procedure is adopted using quantum probability predictions for the two experiments necessary to apply the inequality to experimental data, a different form results for alternative variables' correlation than that assumed by Bell. The inequality is satisfied. This is accomplished without non-locality or non-reality assumptions.

Finally the Bell cosine correlation is derived from a computational model employing three random processes, two Gaussian and one Poisson. The Bell cosine correlation is thus not a unique result of entanglement.

Any who, in spite of the above, believe that the Bell inequalities may in principle be violated by physical data need to perform the following task: write three very small data sets (or four in the four variable case) on the back of an envelope that violate the inequality and show them to the world.

## Acknowledgement

The presentation of the material given above has been influenced by many discussions of the issues with Joe Foremen, personal communications with Karl Hess, and critiques of earlier discussions of this topic by Michael Hall.9

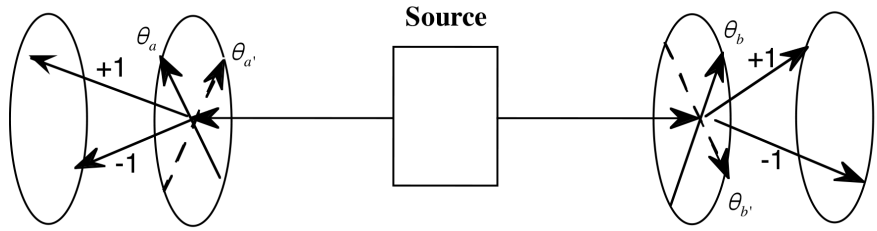

**Figure** 1. Schematic of Bell experiment in which a source sends two particles (photons most often used) to two detectors having angular settings $\theta_a$ and $\theta_b$, (denoted as *a* and *b* in Bell's notation) and alternative settings $\theta_{a'}$ and $\theta_{b'}$. While one measurement operation on the *A*-side, e.g. at setting $\theta_a$, commutes with one on the *B*-side at $\theta_b$, any additional measurements at either $\theta_{a'}$ or $\theta_{b'}$ are non-commutative with prior measurements at $\theta_a$ and $\theta_b$, respectively. This figure was drawn by the author. and modified in notation for use in Ref. 4, as well as other papers.